\pdfoutput=1
\documentclass[a4paper]{article}

\usepackage{amsmath,amssymb,amsfonts,marvosym}
\usepackage{graphicx} %\includegraphics

\DeclareMathOperator\RealPart{{\mathrm{Real}}}

\newcommand{\pd}[2]{\frac{\partial #1}{\partial #2}}
\newcommand{\pdd}[2]{\frac{\partial^2 #1}{\partial #2^2}}
\newcommand{\der}[2]{\frac{d #1}{d #2}}
\newcommand{\derder}[2]{\frac{d^2 #1}{d #2^2}}
\newcommand{\derderder}[2]{\frac{d^3 #1}{d #2^3}}
\newcommand{\cout}[1]{}

\begin{document}

\title{Drag reduction by a solid wall emulating spanwise oscillations. Part 1.}

\author{
S.Chernyshenko\\
 Imperial College London\\
}
\maketitle

\begin{abstract}
A new idea for turbulent skin-friction reduction is proposed, wherein the shape of the solid wall is designed to create the spanwise pressure gradient acting similarly to the well-known method of drag reduction by in-plane spanwise wall motion. Estimates based on the assumption of similarity with drag reduction effect of in-plane wall motion suggest that drag reduction of about 2.4\% can be achieved in the flow past a wavy wall, with  the crests forming an angle of about 38$^\circ$  with the main flow direction, and the wave period in the main flow direction equal to about 1500 wall units. The required height of the wall waves have to be adjusted to achieve the same intensity of the spanwise motion as that created by an in-plane moving wall of the same wavelength and with peak wall velocity equal to 2 wall units. Further research is being conducted in order to determine this height. Suggestions are made for further research on confirming the feasibility of the proposed method and on optimising the wall shape.   
\end{abstract}

\section{Introduction}

\begin{figure}[t]
\centerline{\includegraphics[width=0.35\textwidth]{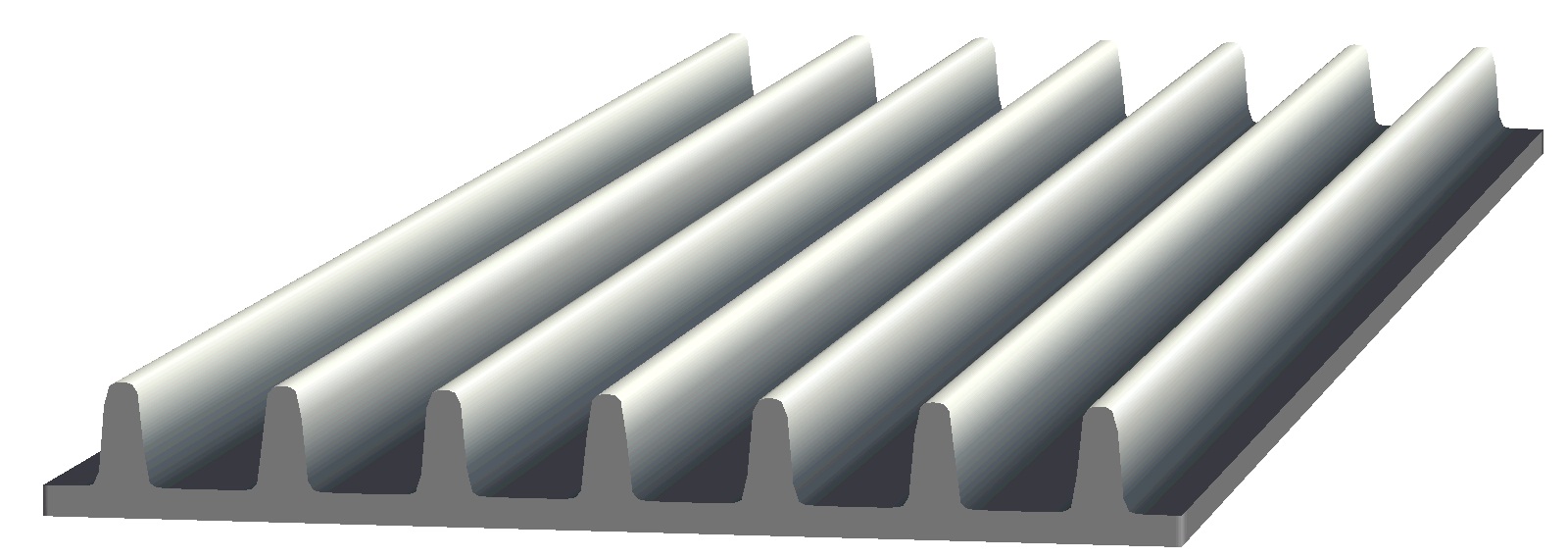}\ \ \includegraphics[clip,trim=6em 2em 5em 4em,width=0.55\textwidth]{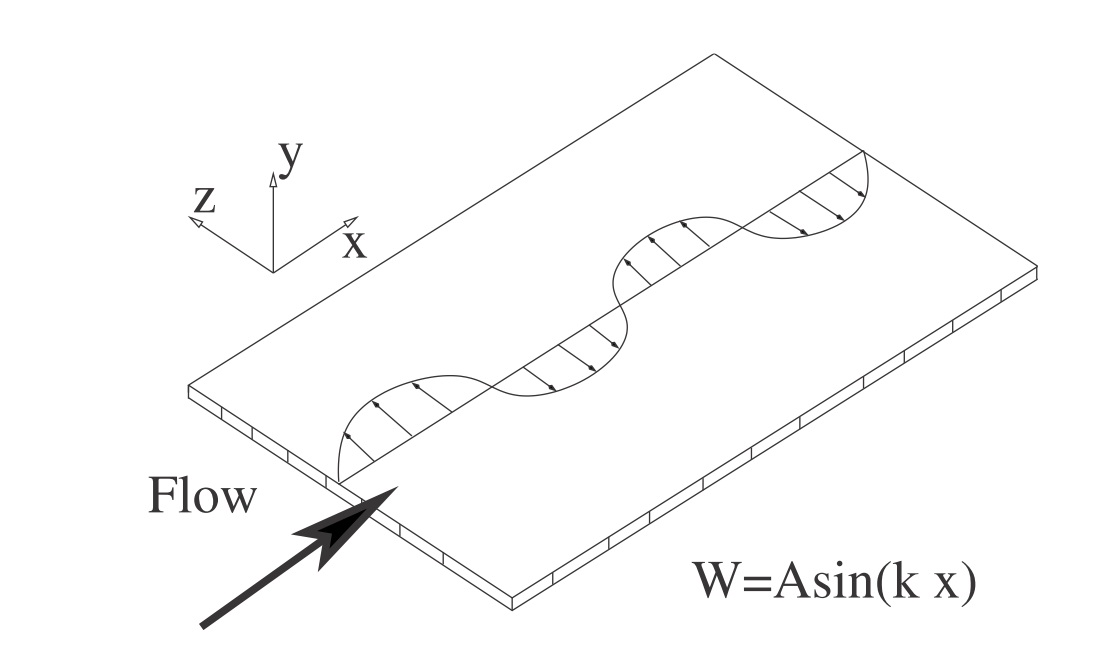}}
\centerline{a)      \hspace{0.5\textwidth}  b)}
\caption{
a)  Riblets;  b) Spanvise wall motion. \label{fig:RibletsSpanwiseWallMotion}}
\end{figure}

Turbulent skin-friction control techniques are classified into passive and active. Passive techniques do not require energy input.  For example, the shape of the solid surface can be made such that the skin friction will be less in the flow past this surface than the skin friction in the flow past a flat wall at the same conditions. The only surface shape known, reliably, to decrease drag is one covered with riblets~\cite{GarciaMayoralJimenez2011}. Riblets are longitudinal grooves in the surface exemplified in Figure~\ref{fig:RibletsSpanwiseWallMotion}a. Riblets inhibit spanwise velocity fluctuations, that are fluctuations directed across the grooves, and this modification of the structure of turbulence reduces the drag. The cross-sectional shape of the riblets can be different from that shown. Importantly, to reduce the drag, both the distance between the neighbouring groves and their height should be about 15 wall units. A wall unit is defined as the distance normalised by the wall skin friction and fluid viscosity.  For aircraft applications a wall unit might be of order of one micron, and riblets have, therefore, to be very small. The drag-reduction level achievable with riblets is less than 12\%, and the extra manufacturing and maintenance costs involved make riblets only marginally effective in a practical environment.

Active control techniques require energy input. For example, the surface of the wall can perform oscillatory in-plane motion in such a way that the turbulent skin-friction drag is reduced. 
Spanwise oscillations of the wall can reduce the skin-friction drag by up to 40\%~\cite{JungMangiavacchiAkhavan1992}. This effect has been predicted computationally and confirmed experimentally \cite{Choi_Clayton_Debisschop_1998}. More complicated in-plane motions, in which different parts of the wall surface move with different time-dependent velocities, can produce even higher effects as was shown in \cite{Quadrio2009}. Importantly, the energy required for such an actuation need only be about half of the gain obtained due to the drag reduction, so that the energy budget is favourable. Of particular interest for the present work are the findings of \cite{Viotti_Quadrio_Luchini_2009}, where a steady wall motion, illustrated in Figure\,\ref{fig:RibletsSpanwiseWallMotion}b, was considered. The authors showed that net energy savings of 23\% were possible. They also found that the optimal longitudinal wavelength of the forcing is somewhat larger than 1000 wall units.

Other approaches to turbulent-flow control, such as blowing and suction, using micro-electro-mechanical actuators with feedback control, or using plasma actuators for creating the cross-flow motion, have been proposed, but so far none of these proposals has resulted in practical applications. 
With the above described approaches not being widely used in practice, there remains a need for a practical and cost-effective control system allowing large skin-friction reductions to be achieved. The present work proposes a simple and practical method of passive control intended to achieve the same effect as the spanwise wall motion shown in Fig.~\ref{fig:RibletsSpanwiseWallMotion}b.

\section{The idea}

\begin{figure}
\centerline{\includegraphics[width=0.51\textwidth]{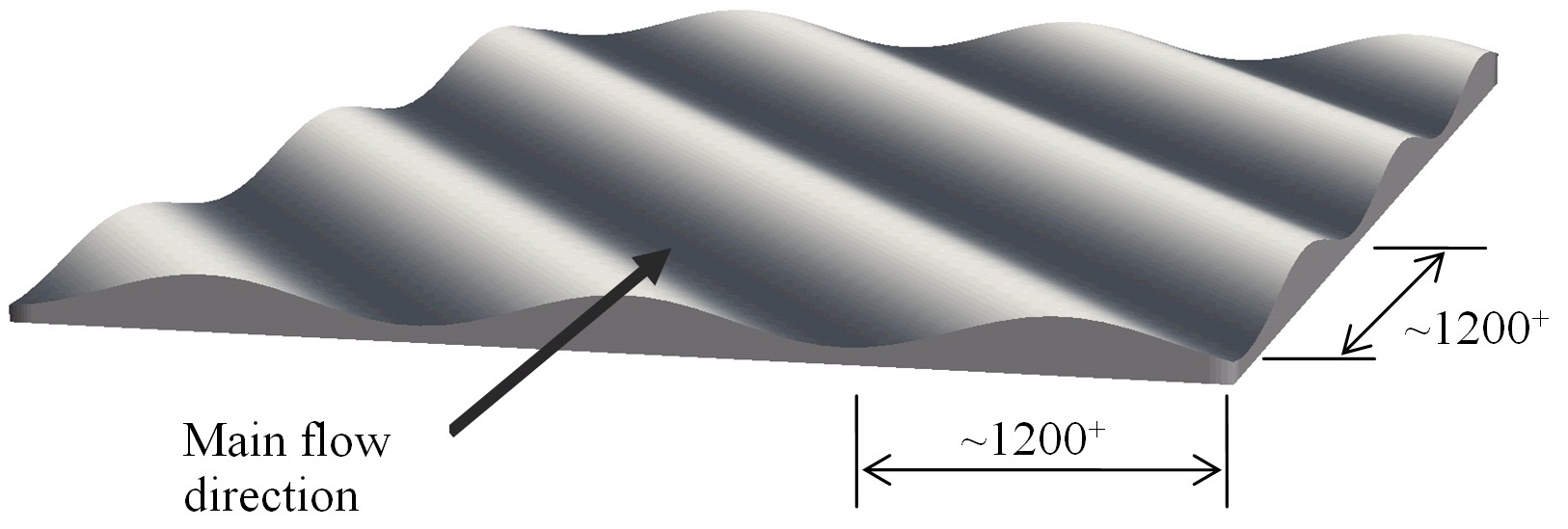}}
\caption{
 Drag-reducing wavy wall\label{fig:WavyWall}}
\end{figure}

The idea consists in generating, by selecting the appropriate shape of the surface, the cross-flow motion producing the drag-reduction effect.  An example of such a wall shape is shown in
Figure\,\ref{fig:WavyWall}. The deflection of the main flow by the wall creates different spanwise pressure gradients on the upwind and downwind sides of the smooth waves. This pressure gradient pushes the fluid sideways, similar to a spanwise wall velocity shown in Figure\,\ref{fig:RibletsSpanwiseWallMotion}b. 

Unlike the case of active control, the rigid wall will not require energy input. However, the additional motion generated by the non-flat wall will result in an increase in energy dissipation similar to that for a moving wall. This additional energy dissipation will manifest itself as a pressure drag on the non-flat surface. Therefore, the proposed method can be considered as a simple and feasible approach to providing energy for generating the drag-reducing spanwise motions.

Crucially, even a relatively small variation of the wall shape will produce a significant variation in the velocity near the wall, where the drag-reduction mechanism is concentrated, because of the mechanism involved in the well-known phenomenon of viscous-inviscid interaction (see, for example, \cite{Messiter1983}).  The properties of boundary layer flow are such that the displacement of streamlines caused by variation in wall surface shape is passed with little change to the streamlines at the outer edge of the boundary layer, where the flow velocity is large. This displacement then results in the pressure variation proportional to the velocity at the outer edge of the boundary layer and to the displacement magnitude. The pressure variation is then passed back to the wall, since the pressure does not vary significantly across the boundary layer. Near the wall this pressure variation results in significantly larger variation of the velocity, because the velocity itself is small near the wall. At the intuitive level this can be understood by recollecting the well-known Bernoulli equation, from which it follows that the pressure variation and the variation of velocity squared are of the same order of magnitude. At the outer edge of the boundary layer the velocity variation $\Delta U,$ created by a small displacement of the streamlines, is small as compared to the velocity $U$ itself, so that the pressure variation $\Delta p$ is of order $U\Delta U,$ but near the wall, where the velocity approaches zero, the velocity variation $\Delta u$ is of the same order as the velocity itself, so that the same $\Delta p$ corresponds to $\Delta u^2.$  Hence, $\Delta u$ is of the same order as $(U/\Delta u)\Delta U,$ that is much greater than $\Delta U.$ 
 For this effect to take place, the characteristic thickness of the near-wall layer should be small as compared to the longitudinal and spanwise length scale of the wall-shape variation. Fortunately, this is possible since the characteristic scale of the near-wall layer is about 100 wall units, as evidenced by the typical streak spacing, while the optimal longitudinal wavelength of the steady spanwise wall motion (Figure\,\ref{fig:RibletsSpanwiseWallMotion}) is somewhat greater than 1000 wall units --- that is, one order of magnitude greater. This wavelength is almost two orders of magnitude greater than the wavelength of riblets. Hence, it is much easier to manufacture and maintain in practice.

%\section {An estimate of the achievable magnitude of drag reduction}

Direct numerical simulation of a flow past such a wall is more computationally expensive than a simulation of a flow past a flat wall, not only because of a more complicated geometry, but also because the waviness of the wall adds an additional spanwise length scale, which is about an order of magnitude greater than the streak spacing. Before performing expensive direct numerical simulations one can attempt to use the similarity between the action of spanwise pressure gradient and the action of spanwise wall motion in order to determine the range of the parameters where the drag reduction is most likely to be observed. This similarity was already reported in \cite{JungMangiavacchiAkhavan1992} where drag reduction by wall oscillations was first demonstrated. In the case considered in that paper the similarity was in fact exact, since the two situations could be made identical by a simple change of the frame of reference. In our case this can only be approximate. We will assume that if the spanwise motion generated by a particular wavy wall is close to the spanwise motion generated by a particular in-plane spanwise wall motion then the effect of this two methods of control will be similar, resulting in the same reduction in the skin friction at the wall. The difference between the two methods of control is that the spanwise motion of the wall consumes power, while a rigid wavy wall does not consume power. However, the wavy wall will experience pressure drag, not present in the flow past a spanwise moving wall, and overtaking this additional drag will require additional power. Thus, in both cases there is a price to be paid for control, which can be expressed as the power required for control. Viotti, Quadrio, and Luchini~\cite{Viotti_Quadrio_Luchini_2009} calculated both the power saved, $P_{\mathrm{sav}}$ and the (negative) power required, $P_{\mathrm{req}},$ for the spanwise-moving wall, and considered their difference, $P_{\mathrm{net}},$ which characterised the net gain achievable by the method in principle. We will find the shapes of the wavy walls generating a spanwise shear similar to the spanwise shear obtained in several cases by \cite{Viotti_Quadrio_Luchini_2009}, and assume that $P_{\mathrm{sav}}$ is the same in both cases.  We will then calculate $P_{\mathrm{req}}$ for a flow past a wavy wall, using an approach demonstrated in \cite{Viotti_Quadrio_Luchini_2009} to work well for the spanwise-moving wall, thus obtaining  $P_{\mathrm{net}}$ for the wavy wall. We then optimise it to get an estimate of the drag reduction achievable in the flow past a wavy wall and the optimal parameters of the wavy wall.

\section{Boundary layer on a wavy wall and the power required}
\subsection{Boundary layer equations for the wavy-wall case and the spatial Stokes
layer case}

Similar to \cite{Viotti_Quadrio_Luchini_2009}, the boundary layer equations, linearized around a linear profile will be used. Written in wall units, they have the form

$$
y\pd ux + v =-\pd px + \pdd uy
$$
$$
0=-\pd py
$$
$$
y\pd wx = - \pd pz + \pdd wy
$$
$$
\pd ux + \pd vy + \pd wz =0
$$

\subsubsection{The spatial Stokes
layer case}

\newcommand\SSL{\mathrm{ssl}}

In the spatial-Stokes-layer (SSL) case the flow is driven by the movement of the boundary, so that  $u=v=0,$ $w=\hat W e^{ik_x x}$ at $y=0.$  The pressure gradient is zero, and $u_{\SSL}=v_{\SSL}=0.$ Taking $w =\hat W \tilde w_{\SSL}(\tilde y) e^{ik_x x},$ where  $y=k_x^{-1/3}\tilde y$ gives  

\begin{equation}\label{eqn:TildeWssl}
i\tilde y\tilde w_{\SSL} = \derder{\tilde w_{\SSL}}{\tilde y}
\end{equation}
with boundary conditions $\tilde w_{\SSL} =1$ at $\tilde y=0$, and $\tilde w_{\SSL}\to0$ as $\tilde y\to\infty,$
the same as in \cite{Viotti_Quadrio_Luchini_2009}. We solved (\ref{eqn:TildeWssl}) numerically and obtained a perfect agreement with the formula  $\tilde w_{\SSL}(\tilde y)=Ai(-i\tilde y e ^{-4\pi i/4})/Ai(0),$ equivalent to (7)  in \cite{Viotti_Quadrio_Luchini_2009}.  

\subsubsection{Wavy wall case}

In the wavy wall case the boundary condition is $u=v=w=0$ at $y=0.$
The pressure distribution is generated by the inviscid flow above the boundary layer. We assume it to have the form
$
p=\hat p e^{i(k_x x + k_z z)}.
$
Accordingly, $(u,v,w)=(\hat u(y), \hat v(y), \hat w(y))e^{i(k_x x + k_z z)}.$
This leads to
\begin{equation}\label{eqn:hatu}
i k_x y \hat u + \hat v = -i  k_x \hat p + \hat u''
\end{equation}
$$i k_x y \hat w  = - i k_z \hat p + \hat w''$$
$$i k_x \hat u +\hat v' + i k_z \hat w =0$$
Eliminating $\hat v$ and then rescaling as 
$$
\hat w(y)=i k_z k_x^{-2/3}\hat p \tilde w(\tilde y),\quad
\hat u(y)=i k_z^2k_x^{-5/3} \hat p \tilde u(\tilde y)
$$
gives, after simple transformations, the following system
$$
i\tilde y \der {\tilde u}{\tilde y}-i\tilde w=\derderder {\tilde u}{\tilde y}
$$
$$
i\tilde y\tilde w = -1 +\derder{\tilde w}{\tilde y}
$$
with boundary conditions $\pd{\tilde u}y\to0$ and $\tilde w\to0$ as $\tilde y\to\infty,$ and $\tilde u = \tilde w =0,$ ${\tilde u}''=k_x^2/k_z^2$ at $\tilde y=0.$ The last condition is in fact (\ref{eqn:hatu}) taken at the wall. The equation for $\tilde w$ separates from the equation for $\tilde u$ and can be solved first. It is then natural to take
$$
\tilde u =\tilde u_w+\frac{k_x^2}{k_z^2}\tilde u_p,
$$
 with  $\tilde u_w$ and $\tilde u_p$ satisfying
$$
i\tilde y \der {\tilde u_w}{\tilde y} - i\tilde w = \derderder {\tilde u_w}{\tilde y},\quad
\tilde u_w\to0 \mathrm{\ as\ } \tilde y\to\infty, \mathrm{\ and\ } \tilde u_w =0, \ {\tilde u_w}''=1 \mathrm{\ at\ } \tilde y=0,
$$
$$
i\tilde y \der {\tilde u_p}{\tilde y}  =\derderder {\tilde u_p}{\tilde y},\quad
\tilde u_p\to0 \mathrm{\ as\ } \tilde y\to\infty, \mathrm{\ and\ } \tilde u_p =0, \ {\tilde u_p}''=0 \mathrm{\ at\ } \tilde y=0.
$$
From the physical viewpoint $\tilde u_w$ corresponds to the perturbation of $u$ due to wall-normal velocity induced by spanwise velocity dependence on $z,$ while $u_p$ is related to the perturbation of $u$ due to the longitudinal pressure gradient induced by the wall.

These ordinary differential equations were solved numerically using Mathematica.

\subsubsection{Matching the spanwise shear profiles in the SSL and wavy-wall case}

\begin{figure}
$\tilde y$

\includegraphics[width=0.15\textwidth]{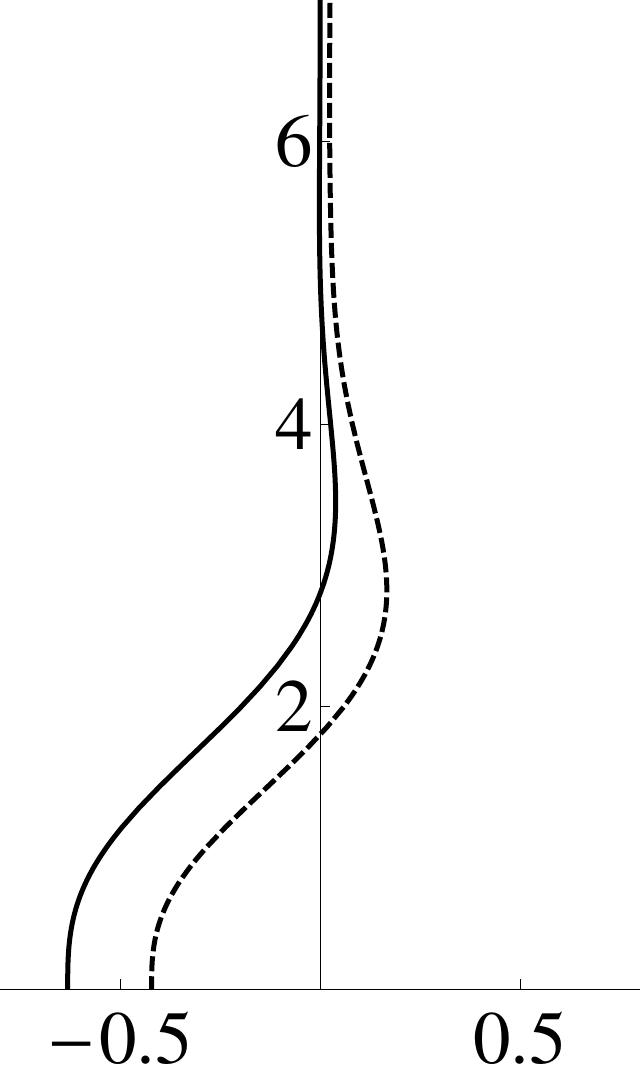}~
\includegraphics[width=0.15\textwidth]{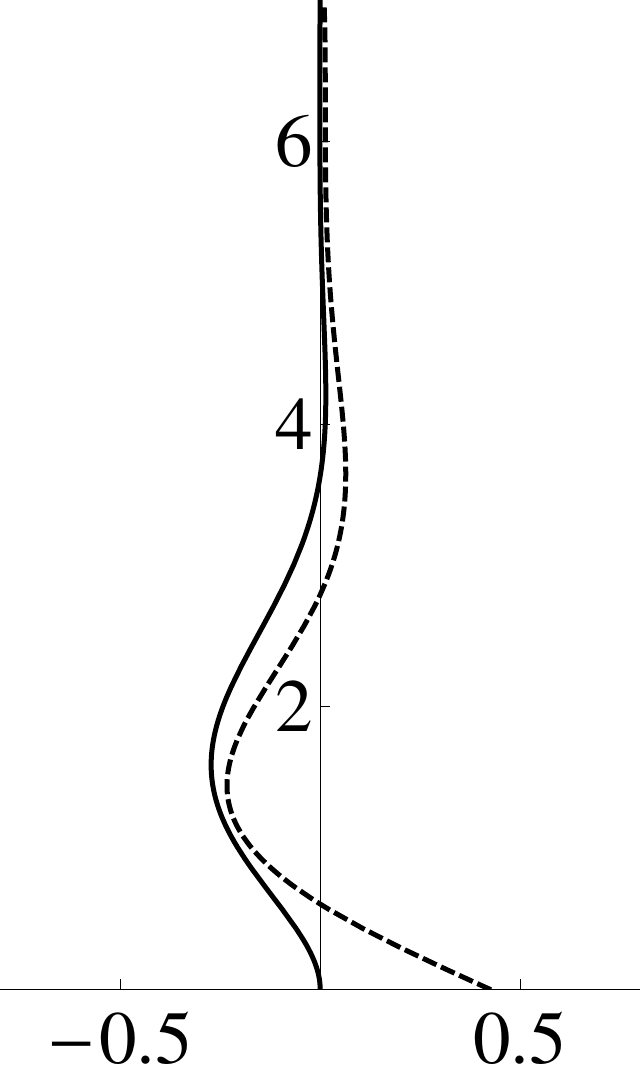}~
\includegraphics[width=0.15\textwidth]{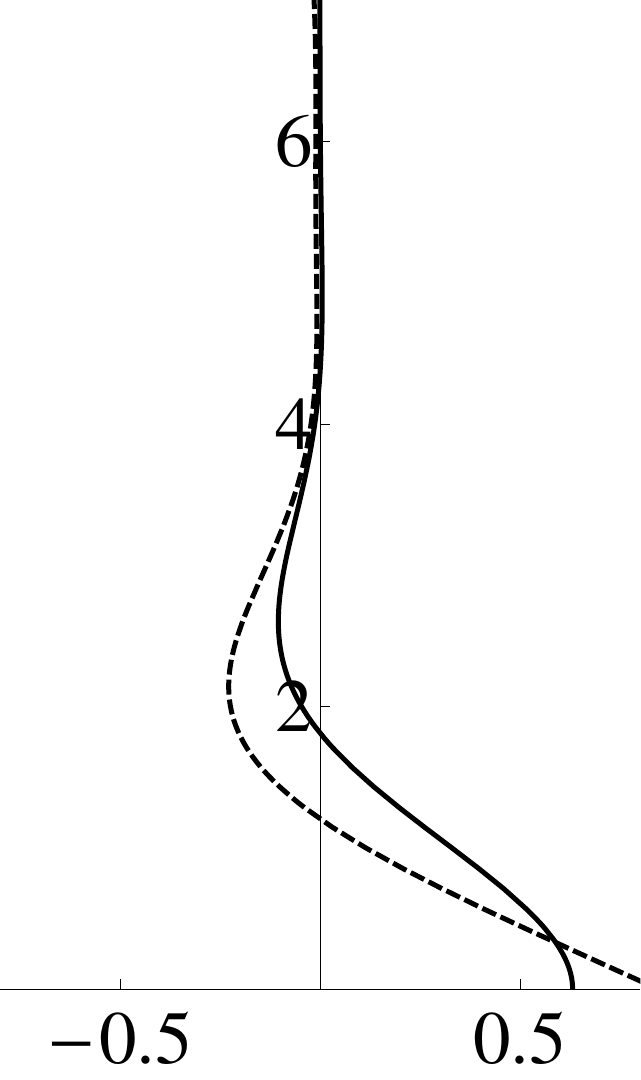}~
\includegraphics[width=0.15\textwidth]{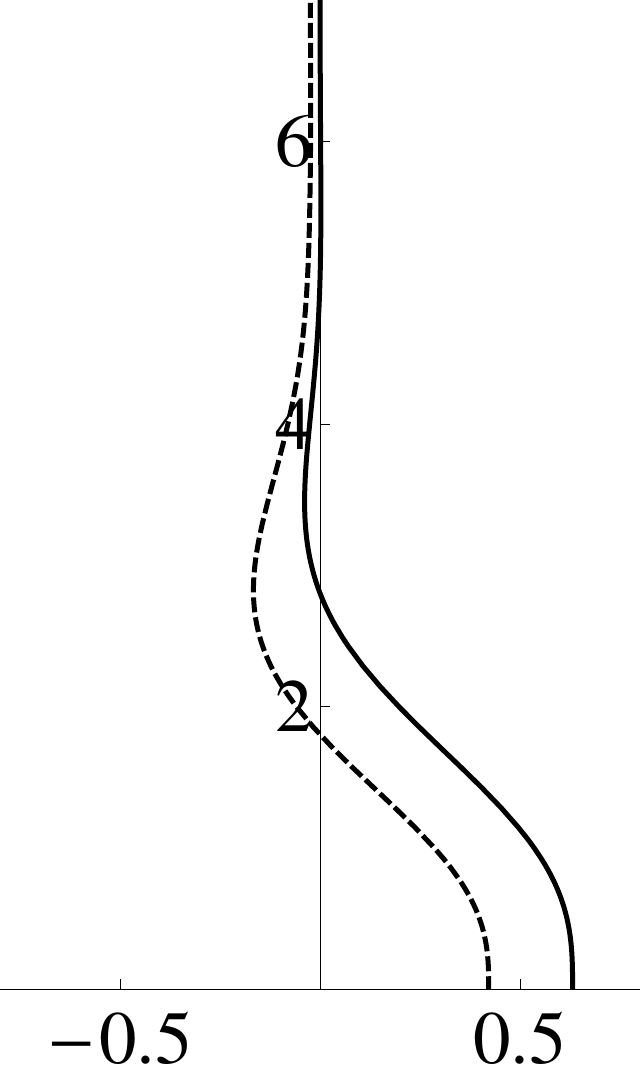}~
\includegraphics[width=0.15\textwidth]{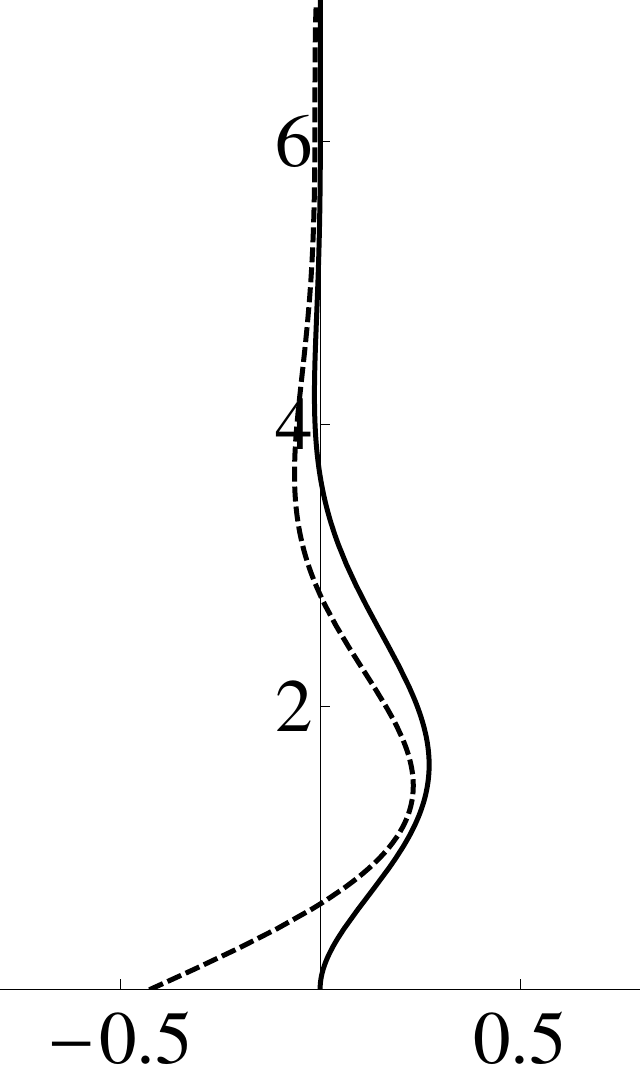}~
\includegraphics[width=0.15\textwidth]{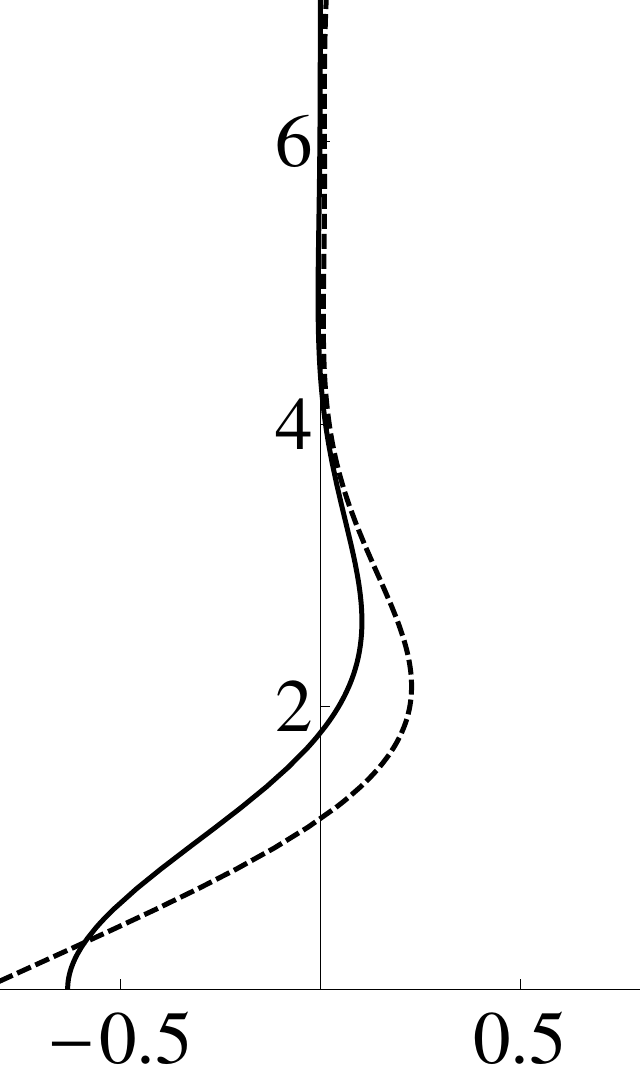}
\caption{Comparison of spanwise shear $\RealPart \tilde w_{\SSL}'(\tilde y)e^{ik_x x}$ (solid) with $\RealPart C_m\tilde w'(\tilde y)e^{i(k_x+\phi_m)}$ (dashed), for $k_x x/(2\pi)=0, 1/6,2/6,3/6,4/6,$ and $5/6.$ 
\label{fig:SpanwiseShear}}   
\end{figure}

At the wall the spanwise velocity is zero in the wavy-wall case and nonzero (and has a maximum amplitude) in the SSL case. Therefore, they cannot be directly matched. However, a simple translational motion might be not so important because of Galilean invariance. We presume that it is the spanwise shear that favourably affects turbulence leading to drag reduction. Therefore, we seek to match the SSL-case spanwise shear  $\hat W \tilde w_{\SSL}'(\tilde y) e^{ik_x x} $ with the wavy-wall-case spanwise shear $i k_z k_x^{-2/3}\hat p \tilde w'(\tilde y)e^{i(k_x x + k_z z)}.$ Note that the wavy-wall period in the spanwise direction is expected to be much larger than the characteristic scale of near-wall turbulent structure and, hence, dependence on $z$ can simply be neglected. On the other hand, a phase shift between the SSL and wavy-wall cases is acceptable. Accordingly, the matching was done by minimising numerically
$$
\int_0^\infty\int_0^{2\pi/k_x} \left|\tilde w_{\SSL}'(\tilde y)e^{i(k_x x)}-\frac{i k_z k_x^{-2/3}\hat p}{\hat W } \tilde w'(\tilde y)e^{i(k_x x + \phi)}\right|^2\,dx\,dy
$$
over $C={i k_z k_x^{-2/3}\hat p}/{\hat W }$ and $\phi.$ This gave $C=C_m=0.8980$ and $\phi=\phi_m=1.5708.$ Hence, to achieve matching, the height of the wall waves should be such that
\begin{equation}\label{eqn:pOfW}
\hat p=C_m{\hat W }/(i k_z k_x^{-2/3})
\end{equation}
Note that $\phi_m\approx\pi/2.$ Figure~\ref{fig:SpanwiseShear} shows the quality of the matching.

\begin{figure}
$\tilde y$

\includegraphics[width=0.15\textwidth]{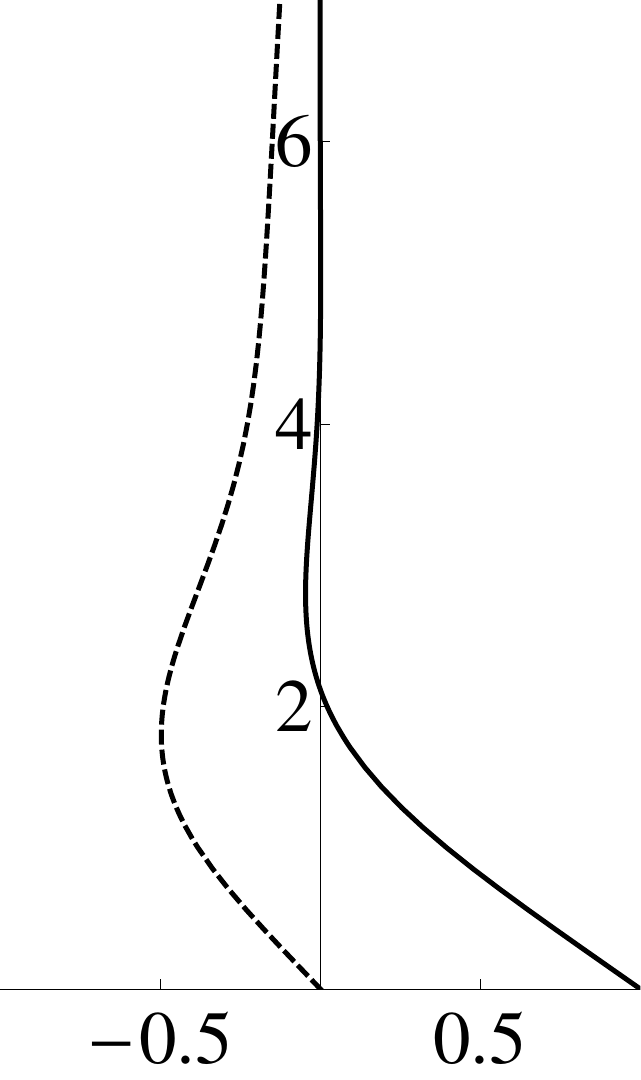}~
\includegraphics[width=0.15\textwidth]{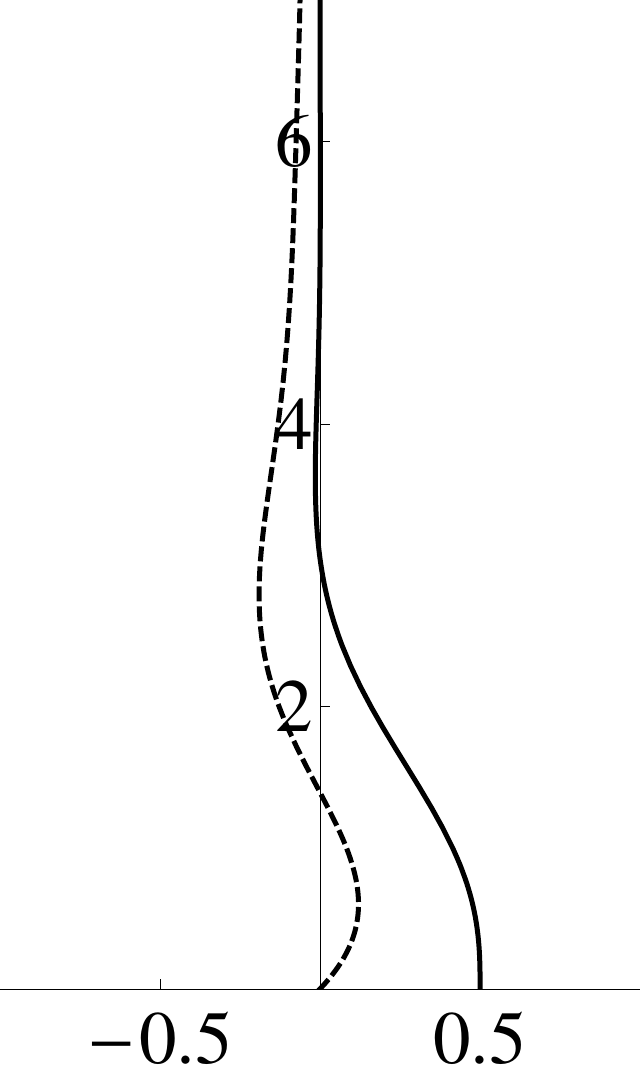}~
\includegraphics[width=0.15\textwidth]{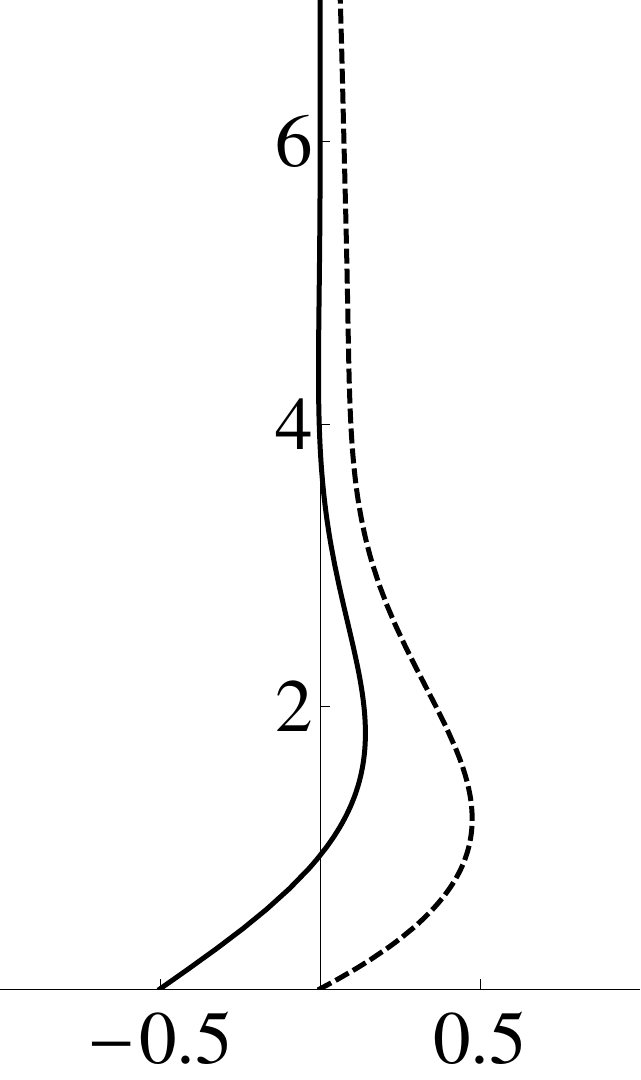}~
\includegraphics[width=0.15\textwidth]{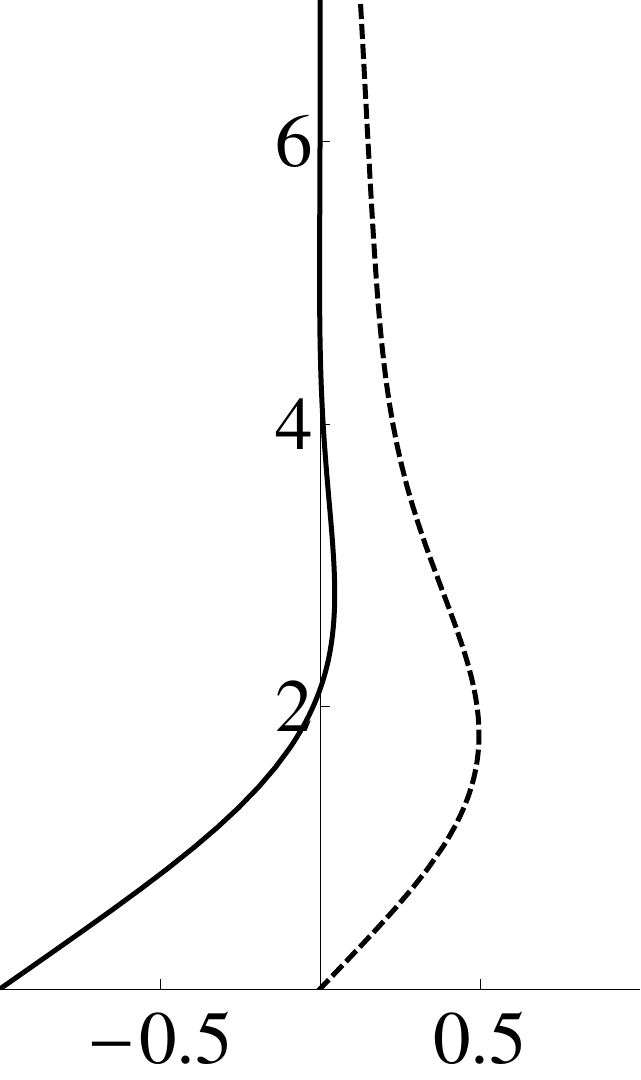}~
\includegraphics[width=0.15\textwidth]{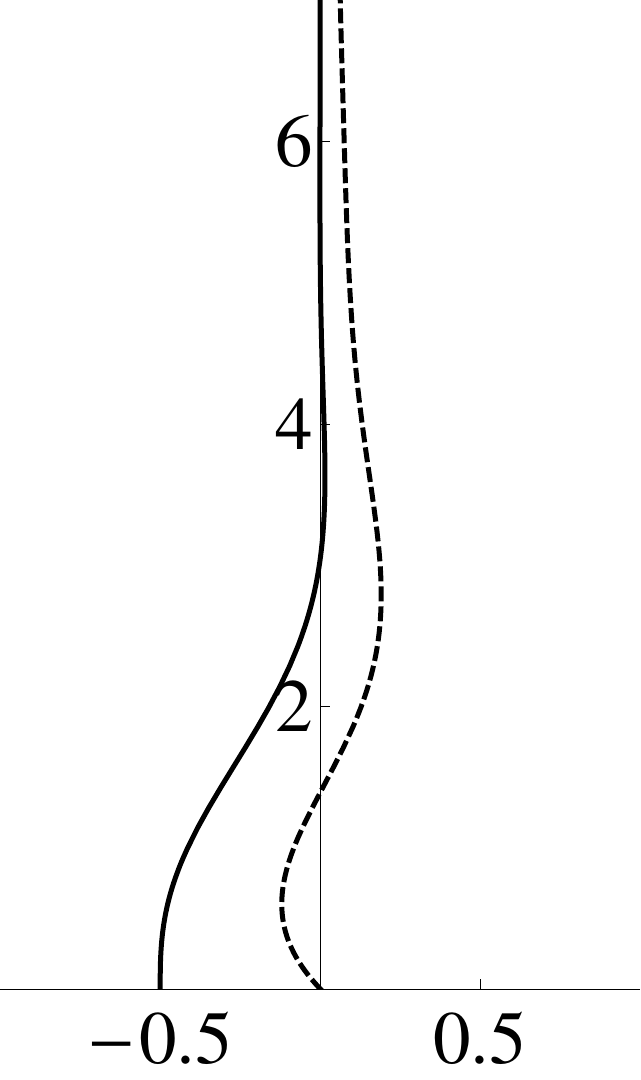}~
\includegraphics[width=0.15\textwidth]{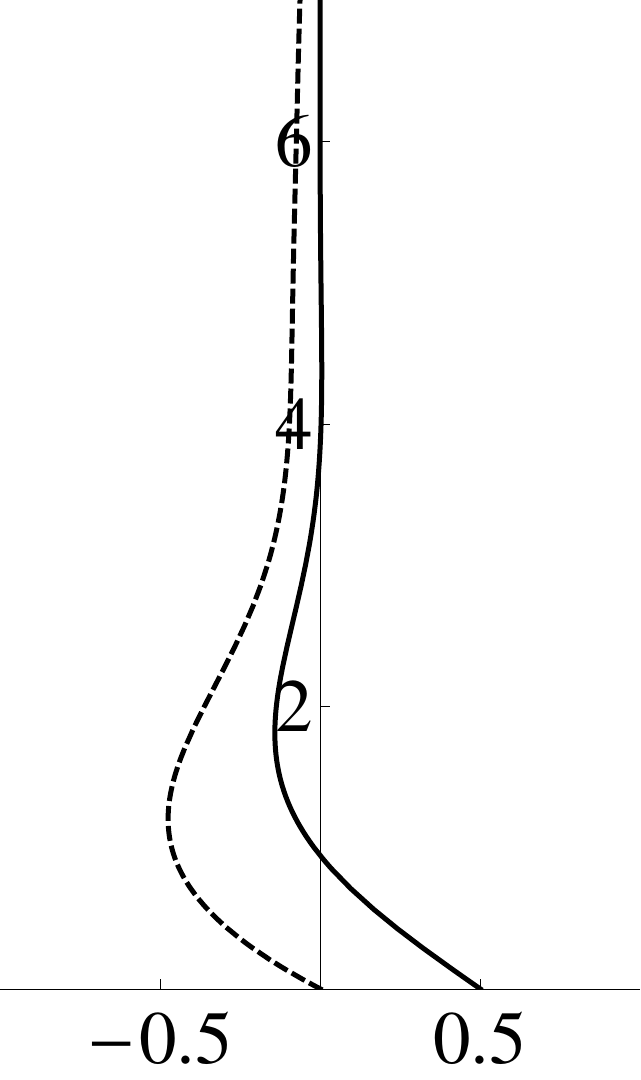}
\caption{Comparison of spanwise velocity $\RealPart \tilde w_{\SSL}(\tilde y)e^{ik_x x}$ (solid) with $\RealPart C_m\tilde w(\tilde y)e^{i(k_x+\phi_m)}$ (dashed), for $k_x x/(2\pi)=0, 1/6,2/6,3/6,4/6,$ and $5/6.$ 
\label{fig:SpanwiseVelocity}}   
\end{figure}

We believe that the agreement is sufficiently close to expect that these two spanwise profiles will lead to similar magnitudes of skin-friction reduction. For completeness, the comparison of the spanwise velocity profiles is shown in Figure~\ref{fig:SpanwiseVelocity}.

 \subsubsection{Comparison of the power required for control in the SSL and wavy-wall cases}

The power per unit wall area (on one wall) required to drive the SSL flow can be expressed in wall units as
 
$$
\Phi^+_{\SSL}=\hat W^2 k_x^{1/3}\int_0^\infty  |  \tilde w'_{\SSL}  |^2/2\,d\tilde y,
 $$
while the corresponding expression in the wavy-wall case is $\Phi^+=\Phi^+_{\mathrm{w}}+\Phi^+_{\mathrm{u}},$ where
$$
\Phi^+_{\mathrm{w}}=  k_z k_x^{-2/3} \hat p \int_0^\infty  |\tilde w'(\tilde y)|^2/2\,d\tilde y,
 $$
$$
\Phi^+_{\mathrm{u}}=  k_z k_x^{-2/3} \hat p \int_0^\infty  |\tilde u'(\tilde y)|^2/2\,d\tilde y.
 $$

If $\hat p$ is selected by the matching rule (\ref{eqn:pOfW}) then $\Phi^+_{\mathrm{w}}\le\Phi^+_{\SSL}.$ This is an artefact of the specific way matching was done: optimising over $C$ is equivalent to projecting the SSL solution onto the direction of the wavy-wall solution in the $L_2$ functional space: the length of the projection of a vector is always less than or equal to the length of the vector itself. To be on the safe side, therefore, it is better to assume that the energy needed to drive the spanwise component of the wavy-wall flow to achieve the same skin friction reduction as in the SSL flow is the same as the energy needed to drive the SSL flow. With this assumption, if the equivalent wavy-wall flow had no $u$ component it would generate the same   overall net drag reduction as the SSL flow. In reality, the $u$ component, however, results in an additional energy dissipation, $\Phi^+_{\mathrm{u}}.$
 Hence, the power required to generate the wavy-wall flow equivalent to the SSL flow will always be greater, with the ratio equal to
$$
r=\frac{\Phi^+}{\Phi_{w}^+}=\frac{||\hat w^2||+||\hat u||^2}{||\hat w^2||}=1+\frac{k_z^2}{k_x^2}\frac{||\tilde u_w+\frac{k_x^2}{k_z^2}\tilde u_p||^2}{||\tilde w ||^2}.
$$
Here, the squared norm $||.||^2=\int_0^\infty |.|^2\,dy.$
The ratio $r$ turns out to be dependent on $k_x/k_z$ only, and it is also obvious that there exist a value of $k_x/k_z$ for which the power required will have a minimum. Calculating the norms numerically gives
$$r=3.122 +2.323
   \frac{k_x^2}{k_z^2}+0.7986\frac{k_z^2}{k_x^2}.
$$

Minimizing $r$ gives $r_{\mathrm{min}}=5.846,$ which is attained at
$$
\left.\frac{k_x}{k_z}\right|_{\mathrm{opt}}=0.7657.
$$ 
 
This corresponds to the angle between the mean flow direction and the direction perpendicular to the wall crests and troughs 
$\Theta_{\mathrm{opt}}=52.56^\circ.$

\subsection{Drag reduction estimate}

\begin{figure}
\begin{center}
\raisebox{0.27\textwidth}{$P_{\mathrm{req}}$}\includegraphics[width=0.5\textwidth]{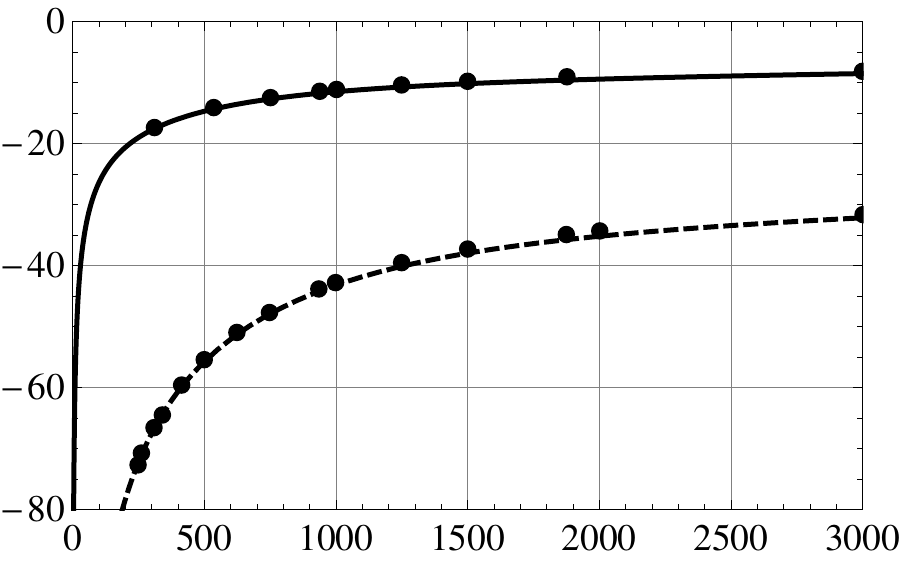}

\ \qquad\qquad\qquad\qquad\qquad\qquad\qquad\qquad$\lambda_x^+$
\end{center}
\caption{Power required to move the wall in SSL flow as a function of the wavelength $\lambda_x^+$; $\hat W^+=6$ (solid), $\hat W^+=12$ (dashed), DNS \cite{Viotti_Quadrio_Luchini_2009} (dots). \label{PowerRequiredSSL}} 
\end{figure}

 In \cite{Viotti_Quadrio_Luchini_2009} the drag reduction was characterised by the net gain in the power needed to drive the flow, expressed as a percentage share of the power needed to drive the uncontrolled flow. The power needed to drive the uncontrolled flow in a plane channel, if expressed in wall units, is
 $$
\Phi^+_0=2U_b^+=2\sqrt{\frac 2{C_f}},$$
 where $U_b^+$ is the bulk velocity. We will use the same empirical formula for the skin friction coefficient $C_f=0.0336{ Re}_\tau^{-0.273}$ as in \cite{Viotti_Quadrio_Luchini_2009}. The net power gain $P_{\mathrm{net}}$ was obtained as a sum of  the power $P_{\mathrm{sav}}$ saved due to the reduction of the skin friction and the (negative) power $P_{\mathrm{req}}$ required to drive the in-plane wall motion. In \cite{Viotti_Quadrio_Luchini_2009} $P_{\mathrm{sav}}$ and $P_{\mathrm{req}}$ were obtained from direct numerical simulations, and expressed in \% of the power needed to drive the uncontrolled flow, and the same convention is used here. It was also shown that with good accuracy $P_{\mathrm{req}}$ can be also obtained from the solution of the linearized boundary layer equation (\ref{eqn:TildeWssl}). It should be understood, of course, that (\ref{eqn:TildeWssl}) is formulated in wall units of the controlled flow while $P_{\mathrm{req}}$ is presented in \cite{Viotti_Quadrio_Luchini_2009} as a function of the wavelength $\lambda^+_x$ in the units normalised with the skin friction of the uncontrolled flow. The reduction of the skin friction is proportional to $100\%-P_{\mathrm{sav}},$ which, after some manipulation, gives the formula
$$
\frac{P_{\mathrm{req}}}{100\%}=-\frac{2\Phi_{\SSL}^+}{\Phi_0^+}=-{\hat W}^2 \sqrt{\frac {C_f}2} \left( \frac{2 \pi(1-P_{\mathrm{sav}}/100\%) }{ \lambda_x^+}\right)^{1/
     3}\int_0^\infty |\tilde w'_{\SSL}(\tilde y)|^2/2\,d\tilde y. $$

This formula, although not given explicitly in \cite{Viotti_Quadrio_Luchini_2009}, was obviously used there.  According to our calculations and the solution given in \cite{Viotti_Quadrio_Luchini_2009}, $\int_0^\infty |\tilde w'_{\SSL}(\tilde y)|^2/2\,d\tilde y=0.3157.$ Comparison of this formula with the DNS results of \cite{Viotti_Quadrio_Luchini_2009}, measured from a digitised plot in that paper, is shown in Figure~\ref{PowerRequiredSSL} and is in fact equivalent to a part of their Figure\,6. The agreement confirms that we use the same approach as in \cite{Viotti_Quadrio_Luchini_2009}.

\begin{figure}
\begin{center}
\raisebox{0.27\textwidth}{$P_{\mathrm{sav}}$}\includegraphics[width=0.5\textwidth]{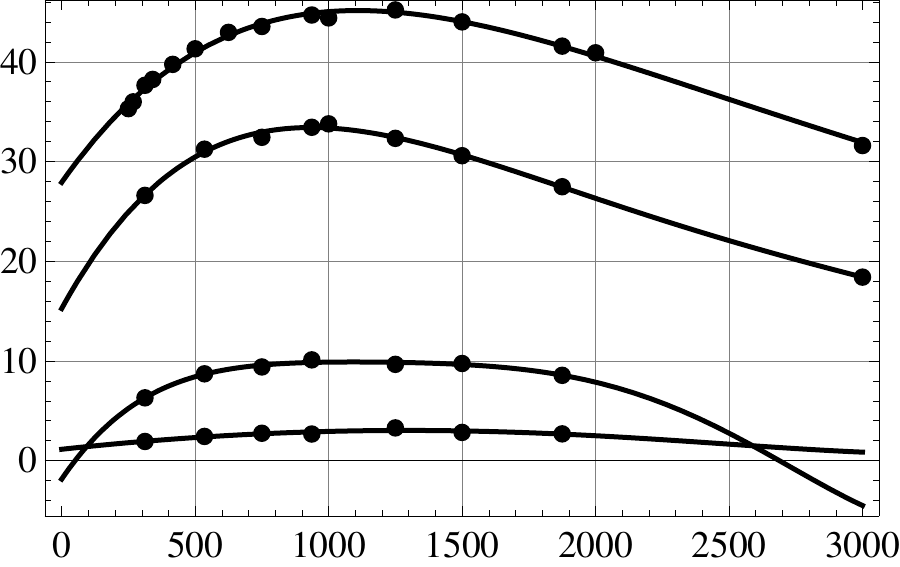}

\ \qquad\qquad\qquad\qquad\qquad\qquad\qquad\qquad$\lambda_x^+$
\end{center}
\caption{
Power saved fits (\ref{eqn:PsavA1}-\ref{eqn:PsavA12}), bottom to top, for the DNS data of   \cite{Viotti_Quadrio_Luchini_2009}.
\label{fig:PowerSaved}
}
\end{figure}
Using the data provided by the authors of \cite{Viotti_Quadrio_Luchini_2009}, the following fits were obtained for $P_{\mathrm{sav}}$ for $\hat W^+=1,2,6$ and $12$:
\begin{multline}\label{eqn:PsavA1}
P_{\mathrm{sav},1}=1.135 + 0.002929 {\lambda_x^+} - 1.205\cdot 10^{-6} {\lambda_x^+}^2 + 
 1.447\cdot 10^{-10} {\lambda_x^+}^3\\ - 1.047\cdot 10^{-13} {\lambda_x^+}^4 + 
 2.609\cdot 10^{-17} {\lambda_x^+}^5,
\end{multline}
\begin{multline}\label{eqn:PsavA2}
P_{\mathrm{sav},2}=-1.856 + 0.03954 {\lambda_x^+} - 5.28537\cdot 10^{-5} {\lambda_x^+}^2 + 
 3.498\cdot 10^{-8} {\lambda_x^+}^3\\ - 1.127\cdot 10^{-11} {\lambda_x^+}^4 + 
 1.328\cdot 10^{-15} {\lambda_x^+}^5,
\end{multline}
\begin{multline}\label{eqn:PsavA6}
P_{\mathrm{sav},6}=15.25 + 0.04888\lambda_x^+ - 4.441\cdot 10^{-5} {\lambda_x^+}^2 + 
 1.628\cdot 10^{-8} {\lambda_x^+}^3\\
 - 2.845\cdot10^{-12} {\lambda_x^+}^4 + 
 1.938\cdot10^{-16} {\lambda_x^+}^5,
\end{multline}
\begin{multline}\label{eqn:PsavA12}
P_{\mathrm{sav},12}=27.90 + 0.03824 {\lambda_x^+} - 2.810\cdot10^{-5} {\lambda_x^+}^2 + 
 8.015\cdot10^-9 {\lambda_x^+}^3\\ - 1.082\cdot10^{-12} {\lambda_x^+}^4 + 
 5.535\cdot10^{-17} {\lambda_x^+}^5.
\end{multline}
The quality of the fits is illustrated by Figure~\ref{fig:PowerSaved}.

\begin{figure}
\begin{center}
\raisebox{0.27\textwidth}{$P_{\mathrm{net}}$}\includegraphics[width=0.5\textwidth]{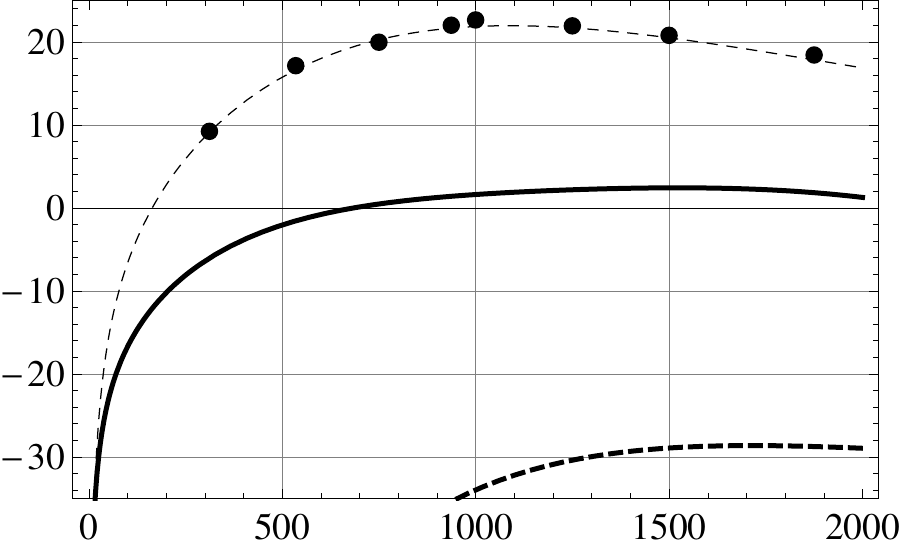}

\ \qquad\qquad\qquad\qquad\qquad\qquad\qquad\qquad$\lambda_x^+$
\end{center}
\caption{
Net power savings for wavy wall case:  $P_{\mathrm{sav},2}+r_m\left.P_{\mathrm{req}}\right|_{\hat W^+=2}$ (solid line) and $P_{\mathrm{sav},6}+r_m\left.P_{\mathrm{req}}\right|_{\hat W^+=6}$ (thick dashed line) and for SSL: DNS  for $\hat W^+=6$ (dots) and   $P_{\mathrm{sav},6}+\left.P_{\mathrm{req}}\right|_{\hat W^+=6}$ (thin dashed line).
\label{fig:DragReduction}
}
\end{figure}

When the wavy wall is used to create the spanwise shear instead of the in-plane wall motion, the power required should be multiplied by the ratio $r$ depending on the angle of the wall wave, with the minimal value $r_m.$ The net power saving is then $P_{\mathrm{sav}}+rP_{\mathrm{req}}.$ Figure~\ref{fig:DragReduction} shows the result for $r=r_m$ for two wall heights equivalent to $\hat W^+=2$ and $6$, together with the $\hat W^+=6$ SSL case. As one can see, in the case corresponding to $\hat W^+=2$ the wavy wall can be expected to give a positive net power saving. The maximum of about 2.4\% drag reduction is attained at $\lambda_x^+\approx1520.$ In the case corresponding to a wavy wall equivalent to SSL with $\hat W^+=6$ only drag increase is predicted. We interpolated between $\hat W^+=1,$ $\hat W^+=2,$ and $\hat W^+=6$ cases, and it turned out that $\hat W^+=2$ gives very nearly the best result.

Note that the calculations were done using the Mathematica package, and while the parameters affecting the accuracy were varied to verify it, the actual accuracy might be somewhat less than four digits given here. In any case, the nature of the present study is such that the values obtained are only indicative. 

\section{Discussion and conclusions}

The present analysis is based on assumptions which, although being reasonable, have not yet been confirmed. The most natural next step will consist in performing direct numerical simulation of a flow past a wavy wall and checking that the drag is indeed reduced. This task is more complicated than a now-standard direct numerical simulation of a flow in a flat channel, not only because of the more complicated geometry, but also because the wavy wall introduces an additional length scale in the spanwise direction, which is noticeably greater than the typical spanwise scale of near-wall structures. For this reason the spanwise dimension of the computational domain might need to be increased. The effect itself is not particularly large, and it is expected to be observed in a relatively narrow range of wall shape parameters. Therefore, a trial-and-error approach to selecting these parameters in direct numerical simulation could be inefficient. Providing a reasonable estimate for these parameters was the main motivation of the present study.

Note that this study is based on the direct numerical simulation results \cite{Viotti_Quadrio_Luchini_2009} for $\mathrm{Re}_\tau=200,$ and the conclusions for other values of the Reynolds number might be somewhat different.

Concerning the required wave height, our results only indicate the magnitude of the spanwise motion that the wall should generate. This might be enough for direct numerical simulations, since estimating approximately this magnitude for a particular wall height does not require averaging over a long time: here, trial-end-error is more likely to work. From the mathematical viewpoint, the required estimate could be obtained by finding the inviscid outer limit of the asymptotic expansion of the flow past a wavy wall as the Reynolds number tends to infinity. This was not undertaken because first direct numerical simulations are likely to be tried for relatively small Reynolds numbers. An alternative is to perform an analogue of the analysis presented in this work numerically using an eddy viscosity model. This line of research is being pursued currently by the colleagues of the author.

Drag reduction by 2.4\% is much smaller than what can be obtained by in-plane wall motion. However, the method proposed here is much easier to implement in practice, since it is passive and does not require wall motion.
It is also easier to implement than riblets, because the wavelength of the proposed wall is two orders of magnitude greater than the wavelength of riblets, and because the height-to-wavelength ratio for the proposed wall is also much smaller than that of riblets.

The present analysis is based on the assumption that similar distributions of spanwise shear will result in similar decrease in turbulent friction, while the longitudinal perturbation of the mean velocity will not affect turbulent friction. This can only be verified in direct numerical simulations or experiment. Note that for the very first tests numerical simulations need not be done for a  wavy wall; the same effect might be expected to be achieved if a steady body force is applied simulating the effect of the pressure distribution caused by a wavy wall.

Another significant source of inaccuracy in the above analysis is the use of linearized boundary layer equations and the assumption that the mean profile in the uncontrolled flow is linear. Therefore, the particular value of 2.4\% should be considered as indication of the accuracy with which the drag needs to be determined in direct numerical simulations rather than a definite prediction. Similarly, the optimal parameters of the wall obtained here are just an indication of the reasonable initial approximation for the search of the true optimal.

It might be possible to increase the effect by optimising the wall shape, since it need not be of the form $\sin{(k_x x + k_z z)}.$ More complicated shapes might be better. Optimisation over complex wall shapes would be impossible on the basis of direct numerical simulations with computing power currently available. Fortunately, recent studies \cite{DuqueDaza2012,Moarref2012} suggest a much more efficient than direct numerical simulation, albeit approximate, approach to such optimisation. Once approximate optimisation using these approaches has been performed, further refinement might be done by direct numerical simulations and experiment.

Since the riblet wavelength and the proposed wall wavelength are so different, they can be combined to work simultaneously. It has already been demonstrated that the effect of wall oscillations and effect of riblets are almost additive~\cite{Vodopianov2013}. 

The main conclusion of the present work can be formulated in the following way. There are reasons to expect that a turbulent flow past a wall the height of which is of the form $h=H\sin (k_x^+ x^+ + k_z^+ z),$ where $x^+$ is the coordinate in the main flow direction, with $k_x^+=2\pi/\lambda_x^+\approx 2\pi/1520$ and $k_x^+/k_z^+\approx 0.7657$ and with a suitable $H$ can be expected to have the skin friction about 2.4\% less than the flow past a flat wall, other things being the same. The value of $H$ should be such as to generate spanwise velocity of the order of $2$ wall units. The wall units here are based on the skin friction in the flow past a flat wall. The values given are only indicative.
\bigskip

This work would be unlikely to appear without the research environment provided by the large team working on the project on turbulent drag reduction under the
 grant EP/G060215/1, funded by EPSRC together with Airbus Operations Ltd and EADS UK Ltd. The author would like to use this opportunity to thank all his colleagues in this team.

%------------------------------------------------------------
%Bibliography
%------------------------------------------------------------
\bibliographystyle{unsrt}
\bibliography{WWbib}

\end{document}